\documentclass[superscriptaddress,floatfix,twocolumn,showpacs,preprintnumbers,amsmath,amssymb,prl]{revtex4}

\usepackage{graphicx}
\usepackage{dcolumn}
\usepackage{bm}

\begin{document}

\preprint{}

\title{Fluctuation-Induced Heat Release from Temperature-Quenched Nuclear Spins near a Quantum Critical Point}

\author{Y. H. Kim}
\affiliation{Department of Physics, University of Florida, P.O.\ Box 118440, Gainesville, Florida 32611-8440,
USA}
\author{N. Kaur}
\affiliation{Department of Chemistry and Biochemistry, Florida State University, Tallahassee, FL 32306, USA}
\author{B. M. Atkins}
\affiliation{Department of Physics, Rhodes College, Memphis, TN 38112, USA}
\author{N. S. Dalal}
\affiliation{Department of Chemistry and Biochemistry, Florida State University, Tallahassee, FL 32306, USA}
\author{Y. Takano}
\affiliation{Department of Physics, University of Florida, P.O.\ Box 118440,
Gainesville, Florida 32611-8440, USA}

\date{\today}

\begin{abstract}
At a quantum critical point (QCP) --- a zero-temperature singularity in which a line of continuous phase
transition terminates --- quantum fluctuations diverge in space and time, leading to exotic phenomena that can
be observed at non-zero temperatures. Using a quantum antiferromagnet, we present calorimetric evidence that
nuclear spins frozen in a high-temperature nonequilibrium state by temperature quenching are annealed by quantum
fluctuations near the QCP. This phenomenon, with readily detectable heat release from the nuclear spins as they
are annealed, serves as an excellent marker of a quantum critical region around the QCP and provides a probe of
the dynamics of the divergent quantum fluctuations.
\end{abstract}


\maketitle

Quantum criticality, the divergence of spatial and temporal extents of quantum fluctuations at a continuous
phase transition at zero temperature, is the driving mechanism behind a variety of novel phenomena that
challenge conventional theoretical approaches to collective behavior of many-body systems and may also lead to
technological applications \cite{Hertz,Sachdev}. This criticality is responsible for the breakdown of
Fermi-liquid behavior in heavy-fermion metals \cite{Lohneysen,Gegenwart} and for the emergence of exotic states
of matter such as unconventional superconductors \cite{Mathur,Yuan}, nematic electron fluids \cite{Fradkin} in
GaAs/AlGaAs heterojunctions \cite{Lilly} and Sr$_3$Ru$_2$O$_7$ \cite{Borzi}, and yet-to-be-identified `reentrant
hidden-order' states in URu$_2$Si$_2$ \cite{Oh}.

Unlike in classical criticality, which is the divergence of order-parameter fluctuations in space at a
continuous phase transition occurring at a non-zero temperature \cite{Domb}, dynamic and static properties are
inextricably linked in quantum criticality \cite{Sondhi,Si}. A spectacular demonstration of this linkage is the
quantum annealing --- the quantum-fluctuation-driven relaxation of quenched disorder --- of an Ising spin glass
near its quantum critical point (QCP), a phenomenon with broad implications in efficient optimization algorithms
\cite{Brooke,Santoro}. Here we report the finding of another relaxation phenomenon near a QCP:
quantum-fluctuation-driven release of heat from nuclear spins. This finding opens up a unique avenue to
investigate the dynamics of quantum fluctuations that underlie quantum criticality.

\begin{figure}[btp]
\begin{center}\leavevmode
\includegraphics[width=0.98\linewidth]{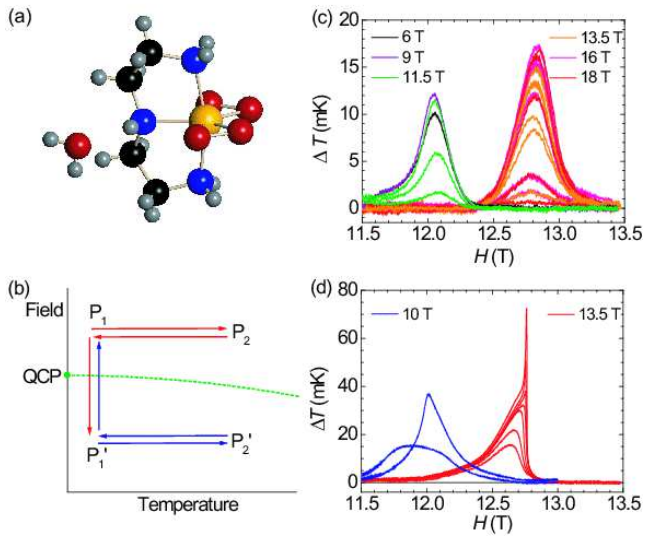}
\caption{(color online). (a) Pair of H$_2$O and Cr(diethylenetriamine)(O$_2$)$_2$ molecules, the basic unit of
the Cr(dien) crystal. The Cr(IV) ion is the largest sphere to the right, bonded to four oxygen and three
nitrogen atoms. (b) Procedure of the experiment (see text). Broken line, terminating in a quantum critical point
at $T=0$, is the boundary between the highly polarized paramagnetic phase and antiferromagnetic phase. (c)
Temperature difference $\Delta T$ between the sample and the thermal reservoir as a function of the magnetic
field during field sweeps at 0.2\,T/min. The thermal reservoir is held at 181\,mK. Temperature $T_q$ from which
the sample has been quenched ranges from 266\,mK to 1.52\,K. (d) Evolution of $\Delta T$ when the thermal
reservoir is held at 96\,mK, as the field is swept at 0.1\,T/min. $T_q$ ranges from 300\,mK to 797\,mK. In (c)
and (d), the peaks to the right are observed during downward field sweeps and those to the left during upward
sweeps. The fields indicated in the legends are $H_q$.}\label{fig1}\end{center}\end{figure}

For this study, we have chosen the inorganic coordination compound
Cr(diethylenetriamine)(O$_2$)$_{2}\cdot$H$_2$O, hereafter referred to as Cr(dien), because of its large number
of hydrogen nuclear spins \cite{House,Ramsey}. This quasi-two-dimensional quantum magnet has a
magnetic-field-tuned QCP at the field $H_c=12.3$\,T, where a highly polarized antiferromagnetic phase gives way
to a field-induced ferromagnetic phase. The Cr(diethylenetriamine)(O$_2$)$_2$ molecule, the key building block
of this material, is an oblate, elongated disk-shaped molecule illustrated in Fig.\ \ref{fig1}(a). The magnetic
ion is Cr(IV), located on the mirror-symmetry plane of the molecule. In the monoclinic crystal structure of
Cr(dien), the $S=1$ spins of Cr(IV) form a square lattice along the crystallographic \emph{ac} plane, with an
exchange energy $J$ of 2.71--2.88\,K.
The spins order antiferromagnetically at
$T_{\mathrm{N}}=2.55$\,K in zero magnetic field \cite{Ramsey}. Application of a high magnetic field depresses
the ordering temperature, pushing it to zero at the critical field $H_c$. We work near this QCP.

The experiment is done in a relaxation calorimeter \cite{Tsujii,calibration}. A 1.02\,mg single crystal of
Cr(dien) is placed on the sample platform, weakly linked to the thermal reservoir through the leads for the
thermometer and heater. This geometry allows the sample temperature to be raised with the heater with ease and
to drop rapidly to the reservoir temperature when the heater is turned off. It also enables the detection of
spontaneously released heat in the sample as an observable temperature difference between the sample and the
reservoir.

The procedure of the experiment is illustrated in Fig.\ \ref{fig1}(b). First, the sample is heated from
temperature $T_0$ (point $P_1$ or $P'_1$ in the figure) of the thermal reservoir to temperature $T_q$ (point
$P_2$ or $P'_2$) ranging from 266\,mK to 1.52\,K, in magnetic field $H_q$ applied perpendicular to the \emph{ac}
plane of the crystal. After 1.4\,min at $T_q$, the sample is rapidly cooled back to $T_0$ in 1.8\,s by turning
off the heater. This temperature quenching leaves the hydrogen nuclear spins frozen in a nonequilibrium
high-energy state corresponding to $T_q$. Subsequently, the magnetic field is swept at 0.2\,T/min or 0.1\,T/min
through the critical field $H_c$. The field sweeps are made at four different $T_0$ ranging from 96\,mK to
261\,mK.

As the swept field approaches $H_c$, heat is released in the sample, manifesting itself as a pronounced peak in
the temperature difference $\Delta T$ between the sample and the thermal reservoir, as shown in Figs.\
\ref{fig1}(c) and \ref{fig1}(d). The heat release occurs only during the first field sweep after the temperature
quenching of the sample from $T_q$, not during subsequent sweeps \cite{magcal}.

\begin{figure}[btp]
\begin{center}\leavevmode
\includegraphics[width=0.98\linewidth]{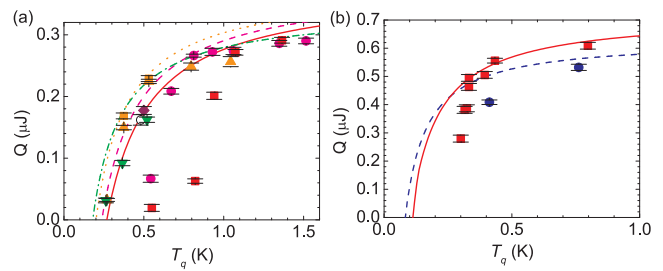}
\caption{(color online). Amount of heat released at (a) 181\,mK and (b) 96\, mK as a function of $T_q$, the
temperature from which the sample has been rapidly quenched. In (a), the sample has been quenched at $H_q=6$\,T
($\circ$), 9\,T ($\blacklozenge$), 11.5\,T ($\blacktriangledown$), 13.5\,T ($\blacktriangle$), 16\,T
($\bullet$), and 18\,T ($\blacksquare$). The lines represent Eq.\ \ref{eq2} with $n_H=10$: dash-dotted line for
$H_q=11.5$\,T, dotted line 13.5\,T, broken line 16\,T, and solid line 18\,T. In (b), the sample has been
quenched at 10\,T ($\bullet$) and 13.5\,T ($\blacksquare$). The lines representing Eq.\ \ref{eq2} are with
$n_H=10$ for $H_q=10$\,T (broken line) and with $n_H=15$, of which five are assumed to freeze only at 132\,mK,
for 13.5\,T (solid line.)}\label{fig2}\end{center}\end{figure}

The amount of released heat $Q$ is obtained from the data via
\begin{equation}
Q=\int\kappa\Delta TdH/\dot{H}, \label{eq1}\end{equation} where $\kappa$ is the thermal conductance of the weak
link between the sample and the thermal reservoir, and $\dot{H}$ the field-sweep rate. As shown in Figs.\
\ref{fig2}(a) and \ref{fig2}(b), $Q$ depends on $T_q$, $H_q$, and $T_0$. These dependences indicate
unambiguously that the heat is indeed released from the hydrogen nuclear spins, as the following analysis
reveals.

When hydrogen nuclear spins in magnetic field $H_q$ are initially frozen in a nonequilibrium high-energy state
determined by temperature $T_q$ --- while the lattice cools rapidly to temperature $T_0$ --- and subsequently
equilibrate with the lattice at another field $H$, the amount of heat released by them is
\begin{equation}
Q=n_{H}nR\left( \frac{\hbar \gamma}{k_{\mathrm{B}}} \right)^2 \frac{I(I+1)}{3} \left(
\frac{H}{T_0}-\frac{H_q}{T_q} \right)H. \label{eq2}\end{equation} Here $n_H$ is the number of hydrogen nuclear
spins (per formula unit) that participate in heat release, $n$ sample's mole number, $R$ the gas constant,
$\hbar$ the Planck constant, $k_B$ the Boltzmann constant, and $\gamma$ and $I = 1/2$ the gyromagnetic ratio and
spin of the hydrogen nucleus. The equation takes into account that the nuclear-spin temperature just before the
heat release is $HT_q/H_q$ instead of $T_q$, as a result of the field sweep from $H_q$ to $H$ being adiabatic
for the nuclear spins except very near $H$.

For $T_0=181$\,mK, good agreement between experiment and Eq.\ \ref{eq2} is obtained with $n_H=10$, as shown in
Fig.\ \ref{fig2}(a), except for a few points for which $H_q$ is 18\,T or 16\,T \cite{wait}. At $T_0=96$\,mK, the
peak that appears at 12.76\,T in Fig.\ \ref{fig1}(d) during downward field sweeps from 13.5\,T is very sharp
while $\Delta T>36$\,mK, \emph{i.e.}, while the sample temperature $T_0+\Delta T$ is higher than 132\,mK. This
indicates that hydrogen nuclear spins whose relaxation times $\tau$ are very short when $T_0>132$\,mK now
participate in heat release. When $T_0=181$\,mK, they have evidently reached thermal equilibrium with the
lattice before each field sweep starts and therefore do not participate in heat release. As shown in Fig.\
\ref{fig2}(b), good agreement between Eq.\ \ref{eq2} and the $T_0=96$\,mK data for $H_q=13.5$\,T is obtained
with $n_H=15$, of which five whose $\tau$ are very short at $T_0>132$\,mK are assumed to freeze at a
nonequilibrium high-energy state at 132\,mK instead of $T_q$. When quenched at $H_q=10$\,T, in the
antiferromagnetic phase, $n_H=10$ gives better agreement with the data, suggesting that those five hydrogen
nuclear spins do not participate in heat release even at $T_0=96$\,mK.

\begin{figure}[btp]
\begin{center}\leavevmode
\includegraphics[width=0.8\linewidth]{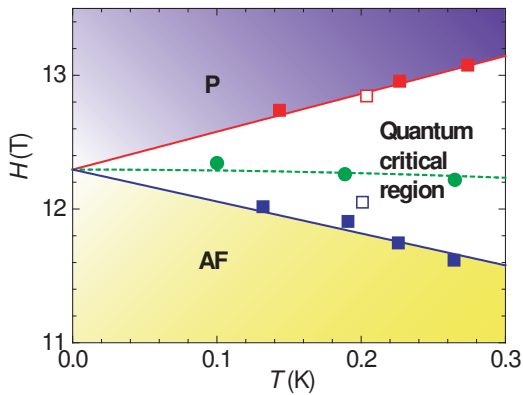}
\caption{(color online). Magnetic fields and temperatures of the peaks in the $\Delta T$ curves (squares),
marking a quantum critical region delimited by two straight lines through the data points. Solid squares are for
peaks at a sweep rate of 0.1\,T/min, open squares at 0.2\,T/min. Circles represent the phase boundary
--- detected by the magnetocaloric effect --- between the antiferromagnetic (AF) and highly polarized
paramagnetic (P) phases, with the broken line from a power-law fit of data points up to 0.84 K. At zero
temperature, the P phase turns into a field-induced ferromagnetic phase.}\label{fig3}\end{center}\end{figure}

\begin{figure}[btp]
\begin{center}\leavevmode
\includegraphics[width=0.8\linewidth]{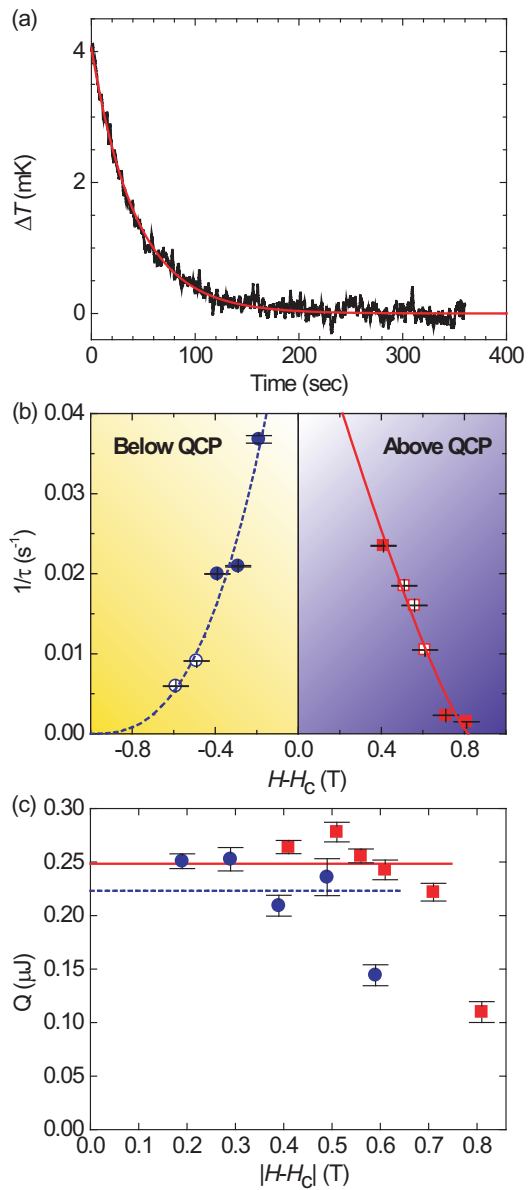}
\caption{(color online). Relaxation of the temperature difference between the sample and the thermal reservoir,
as the magnetic field is brought close to $H_c$ after temperature quenching to $T_0=181$\,mK. (a) Relaxation
curve after a downward field sweep from $H_q=13.5$\,T toward $H_c$ at 0.1\,T/min has been stopped at 12.7\,T.
The sample has been prepared by quenching it from $T_q=796$\,mK at $H_q$. The line is a fit to an exponential
relaxation with $\tau=42.6$\,s. (b) Relaxation rate $1/\tau$ vs. the field at which field sweep has been
stopped. Solid symbols are data from curves showing exponential relaxation, and open symbols from curves showing
stretched-exponential relaxation. Lines are guides to the eye. (c) Amount of heat released vs. the distance from
$H_c$. Horizontal lines indicate the amounts of heat released during complete field sweeps through $H_c$. In (b)
and (c), circles and broken lines are for $T_q=761$\,mK at $H_q=10$\,T, and squares and solid lines for
$T_q=796$\,mK at $H_q=13.5$\,T.} \label{fig4}\end{center}\end{figure}

Cr(dien) contains fifteen hydrogen atoms per formula unit, as shown in Fig.\ \ref{fig1}(a). Among the thirteen
in the Cr(diethylenetriamine)(O$_2$)$_2$ molecule, the five bonded to nitrogens are closer to the Cr(IV) ion
than eight that are bonded to carbons \cite{hydrogen}. It is very likely that the five hydrogens with short
nuclear-spin $\tau$ are those bonded to the nitrogens and thus experience stronger fluctuating dipolar field and
transferred hyperfine field of the Cr(IV) ion, whereas the ten hydrogens with longer $\tau$ are those bonded to
the carbons and the two in the water molecule.

The positions of the heat-release peaks \cite{peaks} are shown in Fig.\ \ref{fig3} along with the phase boundary
between the highly polarized antiferromagnetic phase and similarly highly polarized paramagnetic phase, which
turns into the field-induced ferromagnetic phase at zero temperature, of Cr(dien) near its QCP at $H_c$. This
diagram suggests that, in the zero-temperature limit, the loci of the peaks converge on the QCP. These loci
delimit the region in which $\tau$ is shorter than the timescale of the experiment, a quantum critical region.

How does $\tau$ vary within and near this region? To answer this question, we stop the field sweep at field $H$
and record the subsequent relaxation of the sample temperature toward $T_0$ \cite{T1}. The measurements are
carried out at $T_0=181$\,mK after the sample has been quenched from $T_q=796$\,mK at $H_q=13.5$\,T or from
$T_q=761$\,mK at $H_q=10$\,T. To remove uninteresting contributions coming from the magnetocaloric effect and
eddy-current heating, the measurements are repeated without temperature quenching, and the control data thus
obtained are subtracted from the original data. The relaxation is exponential at some $H$ as shown in Fig.\
\ref{fig4}(a) but shows stretched-exponential behavior at other $H$, with the stretching exponent ranging from
0.61 to 0.78. The relaxation rate $1/\tau$ diverges as the field approaches $H_c$, as shown in Fig.\
\ref{fig4}(b), indicating that diverging quantum fluctuations of the Cr(IV) spins near the QCP drive the
relaxation of the nuclear spins. The divergence is asymmetric around $H_c$, faster for $H>H_c$ than for $H<H_c$.
This asymmetry is also seen in Fig.\ \ref{fig3} as a wider quantum critical region for $H>H_c$ than for $H<H_c$.

The amount of heat released during the relaxation-time measurements is shown as a function of $|H-H_c|$ in Fig.\
\ref{fig4}(c). The result suggests that the ten hydrogen spins that release heat at this temperature contain two
groups, each comprising four to six spins per formula unit. Above $H_c$, one group relaxes at $|H-H_c|$ of about
0.8\,T, whereas the other group relaxes at fields closer to $H_c$. Similarly, below $H_c$, the first group
relaxes at $|H-H_c|$ of about 0.6\,T, whereas the second group relaxes at fields closer to $H_c$. The result is
consistent with the molecular structure of Cr(diethylenetriamine)(O$_2$)$_2$ shown in Fig.\ \ref{fig1}(a): the
eight hydrogens bonded to the four carbons --- with longer nuclear-spin $\tau$ --- fall into two groups, each
consisting of four hydrogens, with distinct ranges of distances from the Cr(IV) \cite{hydrogen}.

In conclusion, our results provide unambiguous evidence that temperature quenching of Cr(dien) leaves the
hydrogen nuclear spins frozen in a nonequilibrium high-energy state and, as the magnetic field is then brought
close to the QCP, quantum fluctuations of the Cr(IV) ionic spins quickly anneal them to reach thermal
equilibrium with the lattice. These results also imply that the quantum-fluctuation-driven heat release from
nuclear spins is a generic phenomenon to be found near a variety of QCP. Because of the inextricable link
between dynamic and static properties in quantum criticality, quantum-critical systems are predicted to exhibit
interesting, non-trivial relaxation phenomena during and after a sweep of a control parameter such as magnetic
field and pressure through the QCP \cite{Patane1} and also after temperature quenching near the QCP
\cite{Patane2}. Our results warn, however, that the response of nuclear spins --- which are nearly ubiquitous
--- to those changing parameters and to quantum fluctuations must be carefully taken into account in real
solids. At the same time, heat release from nuclear spins promises to be a useful probe for the dynamics of
quantum fluctuations that underlie quantum criticality in a variety of systems.

\begin{acknowledgments}
We thank S.\ Nellutla and A.\ Kumar for contributions to an early stage of this work and K.\ Ingersent, P.\
Kumar, Y.\ Lee, D.\ L.\ Maslov, and R.\ Saha for useful discussions. Thanks are also due to J.-H.\ Park, T.\ P.\
Murphy, and G.\ E.\ Jones for assistance and R.\ J.\ Clark for drawing a figure. This work was supported in part
by NSF grant NIRT-DMR-0506946 and by the University of Florida Physics REU program under NSF grant DMR-0552726.
The experiment was performed at the National High Magnetic Field Laboratory, which is supported by NSF
Cooperative Agreement DMR-0654118, by the State of Florida, and by the Department of Energy.
\end{acknowledgments}


\begin{thebibliography}{99}
\bibitem{calibration} The thermometers of the calorimeter have been self-calibrated in magnetic
fields, by measuring the heat capacities of standard materials in situ. In the field range of the present
experiment, calibration uncertainties are about 1.0\% at 0.8 K and about 1.8\% at 0.1 K.
\bibitem{magcal} $\Delta T$ during the subsequent sweeps, exhibiting only the magnetocaloric effect of at most
1\,mK and eddy-current heating by 8\,mK ($T_0=181$\,mK) and 3\,mK ($T_0=96$\,mK), have been subtracted from the
data shown in Figs. 1 and 2.
\bibitem{wait} For those points, 1.4\,min of waiting is probably insufficient to thermalize
the hydrogen nuclear spins.
\bibitem{hydrogen} The five hydrogens bonded to the nitrogens are
located at 2.40--2.52\,{\AA} from the Cr(IV), whereas the eight that are bonded to the carbons fall into two
groups of four, each at 3.08--3.29\,{\AA} and 3.77--3.79\,{\AA} from the Cr(IV). The water hydrogens are at 3.29
and 3.48\,{\AA} from the closest Cr(IV) ions, which are in Cr(diethylenetriamine)(O$_2$)$_2$ molecules not shown
in Fig.\ \ref{fig1}(a). See supporting online information for Ref.\ \cite{Ramsey}.
\bibitem{peaks} The peak position depends slightly on the quenching conditions $T_q$ and $H_q$. For the data
shown in Fig.\ \ref{fig3}, in the highly polarized paramagnetic phase, $H_q=13.5$\,T and $T_q=0.80$\,K except
for the 227\,mK point for which $T_q=1.04$\,K. For those in the antiferromagnetically ordered phase,
$H_q=9$--10.5\,T and $T_q=0.76$\,K except for the 226\,mK point ($T_q=1.00$\,K) and the 201\,mK point taken at
0.2\,T/min ($T_q=0.50$\,K).
\bibitem{T1} $\tau$ measured in this manner is approximately $T_1+C_n/\kappa$ \cite{Andraka},
where $T_1$ is the nuclear spin-lattice relaxation time and $C_n$ the nuclear heat capacity. At 181\,mK,
$C_n/\kappa=$11.1\,s at 12.1\,T.
\end{thebibliography}
\end{document}